\documentstyle[12pt,epsf,epsfig]{article}
\def\ga{\mathrel{\raise.3ex\hbox{$>$\kern-.75em\lower1ex\hbox{$\sim$}}}}
\def\la{\mathrel{\raise.3ex\hbox{$<$\kern-.75em\lower1ex\hbox{$\sim$}}}}
\def\be{\begin{equation}}
\def\ee{\end{equation}}
\def\ba{\begin{eqnarray}}
\def\ea{\end{eqnarray}}
\def\ga{\mathrel{\raise.3ex\hbox{$>$\kern-.75em\lower1ex\hbox{$\sim$}}}}
\def\la{\mathrel{\raise.3ex\hbox{$<$\kern-.75em\lower1ex\hbox{$\sim$}}}}

\newcommand{\bi}[1]{\bibitem{#1}}

\begin{document}

\baselineskip=16pt
\begin{titlepage}
\rightline{UMN--TH--2214/03}
\rightline{FTPI--MINN--03/25}
\rightline{hep-ph/0309252}
\rightline{September 2003}
\begin{center}

\vspace{0.5cm}

\large {\bf
A Re-examination of the $^{187}$Re Bound on the Variation of Fundamental Couplings}
\vspace*{5mm}
\normalsize

{\bf Keith A. Olive}$^{1,2}$, {\bf Maxim Pospelov}$^{3}$, {\bf
Yong-Zhong Qian}$^{2}$,\\  {\bf G\'erard Manh\`es}$^4$,  {\bf Elisabeth Vangioni-Flam}$^5$, 
{\bf Alain Coc}$^6$, and {\bf Michel Cass\'e
}$^{5,7}$

\smallskip
\medskip

$^1${\it Theoretical Physics Institute, School of Physics and
Astronomy,\\  University of Minnesota, Minneapolis, MN 55455, USA}

$^2${\it School of Physics and
Astronomy,\\  University of Minnesota, Minneapolis, MN 55455, USA}

$^3${\it Deptartment of Physics and Astronomy,  University of Victoria, \\
PO Box 3055 STN CSC
 Victoria, BC  V8W 3P6
 CANADA}

$^4${\it Institut de Physique du Globe de Paris,
Laboratoire de G\'eochimie et Cosmochimie, \\
 75252 Paris CEDEX 05 FRANCE}

$^5${\it IAP, CNRS, 98 bis Bd Arago 75014 Paris, FRANCE}

$^6${\it CSNSM, IN2P3/CNRS/UPS, B\^at 104, 91405 Orsay, FRANCE}

$^7${\it SAp, CEA, Orme des Merisiers, 91191 Gif/Yvette CEDEX, FRANCE}

\smallskip
\end{center}
\vskip0.6in

\centerline{\large\bf Abstract}

We reconsider the $^{187}$Re bound on the variation of the
fine-structure constant. We combine the meteoritic measurement
 with several present-day lab measurements to firmly
 establish the quantitative upper limit to the time variation over
 the age of the solar system. We find that the relative change of the 
fine-structure constant between its present value and 
$\alpha$ of $\sim$ 4.6 Gyr ago is 
consistent with zero, $\Delta \alpha/\alpha = [-8\pm 8\  (1\sigma)] \times 10^{-7}$. 
 We also rederive this bound in models where all gauge and Yukawa couplings
vary in an interdependent manner, as would be expected in unified theories.

\vspace*{2mm}

\end{titlepage}


The prospect that the fundamental constants of nature 
vary in time is exciting and represents a radical departure 
from standard model physics. Recent attention to this problem
has focused on the indication that 
the fine structure constant was smaller at 
cosmological redshifts $z=0.5$--3.5 
as suggested by observations of quasar absorption systems 
\cite{webb}--\cite{murphy3}.
The positive result of $\Delta\alpha/\alpha = (0.54 \pm 0.12)\times 10^{-5}$,
where $\Delta \alpha$ is defined as the present value
minus the past one, is based on the many-multiplet method which makes use
 of the $\alpha$ dependence of the relativistic corrections to 
 atomic energy levels.

However, there exist various sensitive experimental checks that
constrain the variation of coupling constants (see e.g., \cite{Sister,uzan}).
Among the most stringent of these is the bound on
 $|\Delta \alpha/\alpha|$ extracted from the analysis
of isotopic abundances associated with the Oklo phenomenon 
\cite{Oklo}--\cite{LV}, a
natural nuclear fission reactor that occurred about 1.8 billion years
ago. While the Oklo bound $|\Delta \alpha/\alpha|< 10^{-7}$ is
considerably tighter than the ``observed" variation, the Oklo phenomenon occurred at a
time period corresponding to $z\approx 0.14$. It is quite
possible that while $\alpha$ varied at higher redshifts, it has not
varied recently.  That is, there is no reason for the variation to be
constant in time. Big bang nucleosynthesis also provides limits on
$\Delta \alpha/\alpha$ \cite{bbn,co,bbn2}. Although these limits are weaker,
they are valid over significantly longer timescales.

Bounds on the variation of the fundamental couplings can also be
obtained from our knowledge of the lifetimes of certain long-lived nuclei.
In particular, it is possible to use relatively precise meteoritic
data to constrain nuclear decay rates back to the time of solar system
formation (about 4.6 Gyr ago). Thus, we can derive a constraint on
possible variations at a redshift $z \simeq 0.45$ bordering the
range ($z = 0.5$--3.5) over which such variations are claimed to be observed.
The pioneering study on the effect of variations of fundamental
constants on radioactive decay
lifetimes was performed by Peebles
and Dicke \cite{PD} and  by Dyson \cite{D}.
The isotopes which are most sensitive to changes in the  $Q$ value
are typically those with the lowest $\beta$-decay $Q$-value, $Q_\beta$. 
The isotope with the smallest $Q_{\beta}$ value ($2.66\pm0.02$~keV) is
$^{187}$Re.

In Dyson's analysis \cite{D}, determination of $^{187}$Re lifetime was 
based on (1) isotopic measurements of molybdenite ores \cite{hirt63},
(2) isotopic measurements of iron meteorites \cite{herr61}, and (3)
a direct measurement of the decay rate \cite{br65}. The decay constant was 
found to be $\lambda = (1.6 \pm  0.2) \times 10^{-11}$,
$(1.4 \pm  0.3) \times 10^{-11}$, and $(1.1 \pm  0.1) \times 10^{-11}$ 
yr$^{-1}$, respectively. The discrepancy between the
direct measurement and the geophysical ones led to speculations that 
this could be due to a time variation in the fine structure 
constant. Note that this would indicate a decreasing value of $\alpha$
from the past to the present, i.e., $\delta \alpha < 0$.
However, the direct measurement \cite{br65} was complicated by many
technical issues, and a later laboratory measurement \cite{dp} gave 
$\lambda = (1.5 \pm  0.2) \times 10^{-11}$ yr$^{-1}$, which was 
consistent with the geophysical measurements.  
As a conservative upper limit, Dyson \cite{D} concluded from the above
measurements that $|\Delta\alpha/\alpha| \le 5\times10^{-6}$
or $|\dot\alpha/\alpha| \le 5\times10^{-15}$ yr$^{-1}$ over the last
Gyr. This was less stringent than the Oklo bound \cite{DD}.

Recently, we have reconsidered the constraint on time variations of the
fundamental couplings based on meteoritic measurements of the lifetimes 
of radioactive isotopes \cite{opqccv}. In particular, we made use of
the dramatic improvement in meteoritic analyses of the 
$^{187}$Re-$^{187}$Os system in iron meteorites. The present abundances 
of $^{187}$Re and $^{187}$Os in a meteorite are
\ba
({^{187}{\rm Re}})&=&({^{187}{\rm Re}})_i\exp(-\lambda t),\label{re}\\
({^{187}{\rm Os}})&=&({^{187}{\rm Os}})_i+
({^{187}{\rm Re}})_i[1-\exp(-\lambda t)],\label{os}
\ea
where the subscript ``$i$'' denotes the initial abundance and $t$ is the
age of the meteorite. Equations (\ref{re}) and (\ref{os}) give
\be
({^{187}{\rm Os}})=({^{187}{\rm Os}})_i+
({^{187}{\rm Re}})[\exp(\lambda t)-1].
\ee
Thus, there is a linear correlation (called an isochron) 
 between the present abundances of
$^{187}$Os and $^{187}$Re
(measured relative to the reference
stable isotope $^{188}$Os, which does not receive any decay contributions). 

The slope of this correlation, $\exp(\lambda t)-1$, can yield the 
$^{187}$Re decay constant $\lambda$ if the age $t$ is independently known.
Radiometric data support 
the idea that the Group IIIA iron meteorites were
 formed at about the same time ($\pm 5$ Myr) as the angrite meteorites  \cite{smo}, which have
 a precisely determined $^{207}$Pb - $^{206}$Pb age of $4.558$ Gyr \cite{lugm}.
The combination of this age with the slope of 
the well defined $^{187}$Re - $^{187}$Os isochron for IIIA iron meteorites gave
  $\lambda=(1.666\pm 0.009)\times 10^{-11}$ yr$^{-1}$ \cite{smo} (see
also \cite{shen,luck}). 
The dominant contribution to the 1$\sigma$ uncertainty is 
associated with
 the non-stochiometry of the Os salt used for calibration ($\approx 0.5\%$).
 Minor uncertainties include the Re - Os isochron slope,
the $^{235}$U and $^{238}$U decay constants \cite{jaf}, the time difference between closure ages for
 U - Pb pair in angrites and the Re - Os pair in IIIAB iron meteorites.

The $\beta$-decay of $^{187}$Re is a unique first fordidden transition,
for which the energy dependence of the decay rate can be approximated as
\cite{RE}
\be
\lambda\propto G_F^2 Q_\beta^3 m_e^2.
\ee
As we showed previously \cite{opqccv}, considering only the variation of
the
Coulomb term in $Q_\beta$, we have
\be
{\Delta\lambda\over\lambda}=3{\Delta Q_\beta\over Q_\beta}\simeq
3\left({20\ {\rm MeV}\over Q_\beta}\right)
\left({\Delta\alpha\over\alpha}\right)\simeq 2\times 10^4
\left({\Delta\alpha\over\alpha}\right).
 \label{la}
\ee
 Thus, given a determination of the possible variation in 
 $\lambda$, we can obtain a limit on the variation of $\alpha$. 
 In our previous work \cite{opqccv}, we assumed that the variation of
$\lambda$ cannot exceed the accuracy of the meteoritic measurements,
$\Delta\lambda/\lambda<0.5\%$. This gave 
$\Delta\alpha/\alpha<3\times 10^{-7}$
or $\dot\alpha/\alpha < 6 \times 10^{-17}$ yr$^{-1}$ 
over a period of 4.6 Gyr, assuming a linear evolution with time.

Two issues concerning the above limit requires attention.
The meteoritic ages used to derive the decay constant of $^{187}$Re
are determined in part by the decay constant of another
radioactive isotope $^{238}$U, which also depends on $\alpha$. However,
as noted earlier \cite{opqccv}, this does not change our limit on
$\Delta\lambda/\lambda$ because the decay constant of $^{187}$Re is far
more (by a factor of 40) sensitive to changes in $\alpha$ than that of 
$^{238}$U. On the other hand, the assumption that the variation
of $\lambda$ cannot exceed the accuracy of the meteoritic measurements
is not justified. Strictly speaking, this assumption is valid if
measurements of comparable precision are made for meteorites with a large
spread (e.g., several Gyr) in age and yield consistent decay constant for
$^{187}$Re. In practice, accurate measurements are available only for
iron meteorites with essentially the same age. Thus, our earlier limit 
based on $\Delta\lambda/\lambda<0.5\%$ over a period of 4.6 Gyr
was over-restrictive.

To obtain a solid limit on $\Delta\lambda/\lambda$, we can compare
the meteoritic measurements of $\lambda$, which cover 4.6 Gyr, with
direct laboratory measurements, which give the present value $\lambda(t_0) = \lambda_0$.
Indeed, there is a reasonably accurate direct measurement which yields 
$\lambda_0 = (1.639 \pm 0.025) \times 10^{-11}$ yr$^{-1}$ 
\cite{lin1} (see also \cite{lin2}). The variation in $\lambda$ is then 
$\Delta\lambda \ =  \lambda_0 - \lambda = (-0.027 \pm 0.026)\times 10^{-11}$
yr$^{-1}$, or
\be
{\Delta \lambda \over \lambda} = -0.016 \pm 0.016.
\label{dlambda}
\ee
As one can see, there is an insignificant trend in the negative direction.
However, it is worth mentioning that the implication of 
the quasar absorption system measurements would be a {\em positive} 
$\Delta \lambda$, under the natural assumption that $\alpha(t)$ is a 
smooth monotonic function.
This result and Eq. (\ref{la}) give $\Delta \alpha/\alpha =
[-8\pm 8\  (1\sigma)]\times 10^{-7}$, which yields a $2\sigma$ limit
\be
-24 \times  10^{-7} < {\Delta \alpha \over \alpha} < 8 \times 10^{-7}.
\label{dalpha}
\ee
This is weaker than the limit reported in \cite{opqccv} by a 
factor of $\sim 3$ for 
a positive $\Delta \lambda$ and by a factor of $\sim 8$ for negative 
$\Delta \lambda$. Even so, the $O(10^{-6})$
limits imposed by the meteoritic data  
still provide strong constraints on 
models of $\alpha(t)$ and severely restrict possibilities to 
accomodate the claimed variation of $\alpha$ based on 
observations of quasar absorption systems (see, e.g. Ref. \cite{cnp}).
This result is also about 6 times stronger than that given in 
\cite{lin2} based on the meteoritic data of \cite{luck}.
(We note that a more recent measurement \cite{RE} 
gave $\lambda_0 = 1.68\times 10^{-11}$ yr$^{-1}$ with a larger systematic 
uncertainty of $\approx 3\%$. Using a weighted average of the two direct
measurements of $\lambda_0$ does not affect the above result significantly.)

It is important to note that this approach, i.e. the comparison of the meteoritic 
measurement of $\lambda$ with $\lambda_0$, will benefit from any improvement in the determination of 
$\lambda_0$ which is required for the application of the Re - Os geochronology \cite{bege}.
On the other hand, it is also important to recognize the limitations of this limit.
Strictly speaking, if $\lambda$ varies in time, then the meteoritic measurement
really represents the time average of $\lambda(t)$ over the age of the meteorite.
As such, the redshift at which the limit can be applied will in principle
depend on the specific functional dependence of $\lambda(t)$.  For example,
if we consider a simple mononomial form, $\lambda(t) - \lambda_0 \sim (t_0 - t)^n$,
then the limits (\ref{dlambda}) and (\ref{dalpha}) should be applied at
a reduced look-back time of $(t_0 - t) = 4.6 {\rm Gyr} / (n+1)^{1/n}$.
As noted above, a look-back time of 4.6 Gyr corresponds to a redshift of 
$z \sim 0.45$ for the assumed cosmological model, hence the limit would
be applied (in a model dependent way) at a somewhat smaller redshift.
For example, in the linear case, $n=1$, one can either apply our limits above at 
a time $t_0 -t = 2.3$ Gyr, or one can apply a relaxed limit (by a factor of 2) 
at a look-back time of 4.6 Gyr.
Because of the model dependence, it may be possible with a careful choice of a
time dependence for $\lambda$ to obviate this limit \cite{fuj} due to the limit being
tied to a time average of the decay rate.

As we discussed previously, in the context of unified or string-inspired 
theories, all Yukawa couplings and gauge couplings depend on the same moduli 
fields and the change in the fine structure constant 
typically implies a change in other couplings and mass scales 
\cite{co,ds}.  The dominant effects
are found in induced variations of the  QCD scale $\Lambda$ and the Higgs 
expectation value $v$. The variations
${\Delta \Lambda \over \Lambda}\simeq 30 {\Delta \alpha \over \alpha}$
and ${\Delta v \over v} \sim 80 {\Delta \alpha \over \alpha}$
\cite{co,lss,other} are translated into
variations in all low energy particle masses.  In short, once we
allow $\alpha$ to vary, virtually all masses and couplings are expected
to vary as well, typically much more strongly than the variation induced
by the Coulomb interaction alone.
 In Ref. \cite{opqccv}, we adopted $|\Delta
\Lambda/\Lambda - \Delta v/v| \sim 50 \Delta \alpha/\alpha$.
The contributions to
$Q_\beta$ from (the kinetic energy) $T$ and (the nuclear potential energy) 
$V$, which scale with $\Lambda$, are comparable to
that from the dominant Coulomb term $(C)$, which scales as $\alpha \Lambda$.
As changes in $\Lambda$ are $O(30)$ times larger than that in $\alpha$,
we can estimate
\be
{\Delta\lambda\over\lambda}=3{\Delta Q_\beta\over Q_\beta} -  2
{\Delta v \over v} \simeq 3{T(V,C)\over
Q_\beta}{\Delta\Lambda\over\Lambda}\simeq
2 \times 10^4{\Delta\Lambda\over\Lambda},
\label{Rhlimit}
\ee
which gives
\be
-8 \times 10^{-8} < {\Delta\alpha\over\alpha}< 3 \times 10^{-8},
\ee
over 
a period of 4.6 Gyr.

In summary, we have revisited the bounds on the change of the 
fine-structure constant from the $^{187}$Re meteoritic measurements.
We have shown that these limits are determined by the accuracy of the 
laboratory determination of the decay rate. This results in a bound, 
$\Delta \alpha/\alpha = (-8\pm 8)\times 10^{-7}$ going back to 
$\sim$ 4.6 Gyr ago, that is of significant importance for all models of $\alpha(t)$. 

\noindent{ {\bf Acknowledgments} } \\
\noindent
We would like to thank I. Lyon for very helpful comments.
This work was supported in part by DOE grants
DE-FG02-94ER-40823, DE-FG02-87ER40328, and DE-FG02-00ER41149
at the University of Minnesota, by NSERC of Canada and by PICS 1076 CNRS
France/USA.



\begin{thebibliography}{99}


\bibitem{webb}
J.~K.~Webb, V.~V.~Flambaum, C.~W.~Churchill, M.~J.~Drinkwater and J.~D.~Barrow,
Phys.\ Rev.\ Lett.\  {\bf 82} (1999) 884
[arXiv:astro-ph/9803165].


\bibitem{murphy1}
M.~T.~Murphy {\it et al.},
Mon.\ Not.\ Roy.\ Astron.\ Soc.\  {\bf 327} (2001) 1208
[arXiv:astro-ph/0012419].
J.~K.~Webb {\it et al.},
Phys.\ Rev.\ Lett.\  {\bf 87} (2001) 091301
[arXiv:astro-ph/0012539].

\bibitem{murphy2}
M.~T.~Murphy, J.~K.~Webb, V.~V.~Flambaum, C.~W.~Churchill and J.~X.~Prochaska,
Mon.\ Not.\ Roy.\ Astron.\ Soc.\  {\bf 327} (2001) 1223
[arXiv:astro-ph/0012420].

\bibitem{murphy3}
M.~T.~Murphy, J.~K.~Webb and V.~V.~Flambaum,
arXiv:astro-ph/0306483.


\bibitem{Sister}
P.~Sisterna and H.~Vucetich,
Phys.\ Rev.\ D {\bf 41} (1990) 1034.

\bibitem{uzan}
J.~P.~Uzan,
Rev.\ Mod.\ Phys.\  {\bf 75} (2003) 403
[arXiv:hep-ph/0205340].


\bi{Oklo} A. I. Shlyakhter, Nature {\bf 264} (1976) 340.

\bi{DD} T.~Damour and F.~Dyson,
Nucl.\ Phys.\ B {\bf 480} (1996) 37.

\bi{Fujii} Y. Fujii {\em et al.}, Nucl. Phys. {\bf B573} (2000) 377.

\bi{LV} S.~J.~Landau and H.~Vucetich,
astro-ph/0005316; N.~Chamoun, S.~J.~Landau and H.~Vucetich,
Phys.\ Lett.\ B {\bf 504} (2001) 1.

\bibitem{bbn} E.~W.~Kolb, M.~J.~Perry and T.~P.~Walker,
Phys.\ Rev.\ D {\bf 33}, 869 (1986);
R.~J.~Scherrer and D.~N.~Spergel,
Phys.\ Rev.\ D {\bf 47}, 4774 (1993);
L.~Bergstrom, S.~Iguri and H.~Rubinstein,
Phys.\ Rev.\ D {\bf 60}, 045005 (1999)
[arXiv:astro-ph/9902157];
K.M. Nollett and R.E. Lopez, astro-ph/0204325.

\bibitem{co}B.~A.~Campbell and K.~A.~Olive,
Phys.\ Lett.\ B {\bf 345}, 429 (1995)
[arXiv:hep-ph/9411272].

\bibitem{bbn2}
K.~Ichikawa and M.~Kawasaki,
arXiv:hep-ph/0203006.


\bibitem{PD} P.J. Peebles and R.H. Dicke, Phys. Rev., {\bf 128}, 2006
(1962).

\bibitem{D} F.J. Dyson, in Aspects of Quantum Theory edited by A. Salam
 and E.P. Wigner (Cambridge U. Press), p. 213 (1972).

\bibitem{hirt63} 
B. Hirt, G. Tilton, W. Herr, and W. Hoffmeister, In Earth Science and Meteoritics, ed. J. Geiss and E.D. Goldberg (North-Holland, Amsterdam, 1963).

\bibitem{herr61}
W. Herr, W. Hoffmeister, B. Hirt, J. Geiss, and F.G. Houtermans, Z. Naturforsch {\bf 16a} (1961) 1053.

\bibitem{br65}
R.L. Brodzinsky and D.C. Conway, Phys. Rev. {\bf 138} (1965) B1368.

\bibitem{dp}
R.W.P. Drever and J.A. Payne, unpublished.

\bibitem{opqccv}
K.~A.~Olive, M.~Pospelov, Y.~Z.~Qian, A.~Coc, M.~Casse and E.~Vangioni-Flam,
Phys.\ Rev.\ D {\bf 66} (2002) 045022
[arXiv:hep-ph/0205269].

\bibitem{smo} M. I. Smoliar et al., Science, {\bf 271}, 1099 (1996).

\bibitem {lugm} G.W. Lugmair and S.J.G. Galer, Geochim. Cosmochimica Acta, {\bf 56}, (1992) 1673.


\bibitem{shen} J. J. Shen, D. A. Papanastassiou, and G. J. Wasserburg,
Geochimica et Cosmochimica Acta {\bf 60}, 2887 (1996).


\bibitem{luck}
J.-M. Luck, J.-L. Brick, and C.-J. Allegre, Nature {\bf 283} (1980) 256;
J.-M. Luck and C.-J. Allegre, Nature {\bf 302} (1983) 130.

\bibitem {jaf} A.H. Jaffey, K.F. Flynn, L.E. Glendenin, W.C. Bentley and
 A.M. Essling, Phys. Rev., C4, (1971) 1889.


\bibitem{RE} M. Galeazzi, F. Fontanelli, F. Gatti, and S. Vitale,
Phys. Rev. C, {\bf 63}, 014302 (2000).



\bibitem{lin1} M. Lindner et al., Geochimica. Cosmochimica. Acta, {\bf
53}, 1597 (1989).

\bibitem{lin2} M. Lindner et al., Nature, {\bf
320} (1986) 246.


\bibitem{cnp} E.~J.~Copeland, N.~J.~Nunes and M.~Pospelov,
arXiv:hep-ph/0307299.


\bibitem{bege} F. Begemann, K.R. Ludwig, G.W. Lugmair, K. Min, L.E. Nyquist, 
 P.J. Patchett, P.R. Renne, C.-Y. Shih, I.M. Villa and R.J. Walker, Geochim Cosmochimica Acta {\bf 65} (2001) 111.

\bibitem{fuj}
Y. Fujii and A. Iwamoto, hep-ph/0309087.



\bibitem{ds} V.V. Dixit and M. Sher, Phys. Rev. {\bf D37} (1988) 1097.





\bibitem{lss}P.~Langacker, G.~Segre and M.~J.~Strassler,
Phys.\ Lett.\ B {\bf 528}, 121 (2002)
[arXiv:hep-ph/0112233].


\bibitem{other}
T.~Dent and M.~Fairbairn,
arXiv:hep-ph/0112279;
X.~Calmet and H.~Fritzsch,
arXiv:hep-ph/0112110;
arXiv:hep-ph/0204258;
T.~Damour, F.~Piazza and G.~Veneziano,
arXiv:gr-qc/0204094;
arXiv:hep-th/0205111.





\end{thebibliography}
\end{document}